\documentclass[11pt]{article}
\usepackage{hyperref}
\usepackage{bm}
\usepackage[numbers,sort&compress]{natbib}
\pdfoutput=1
\begin{document}
\title{Shear-Induced Droplet Breakup and Subsequent Coalescence of the Daughter Droplets}
\author{Orest Shardt, Alexandra Komrakova, J.J. Derksen, Sushanta K. Mitra \\
\\\vspace{6pt} University of Alberta, Edmonton, AB, T6G 2R3, Canada}
\maketitle
\begin{abstract}
In this fluid dynamics video we show simulations of droplet breakup and collisions in simple shear flow. Due to the high resolution of the GPU-based simulations, we capture the transition from coalescence at low capillary numbers to sliding at higher values.
\end{abstract}
\section*{Description}

A liquid drop of radius $R$ is suspended in another liquid. This drop is placed between two parallel plates separated by a distance $H$ and subjected to a simple shear flow. Both plates move with a speed $u_w$ in opposite directions, producing a shear rate of $\dot{\gamma}=\frac{2u_w}{H}$. The liquids have equal densities $\rho$ and viscosities $\mu$. The interfacial tension between the liquids is $\sigma$. The behaviour of the system is determined by viscous, capillary, and inertial forces. The capillary number $\mathrm{Ca} \equiv \frac{\mu\dot{\gamma}R}{\sigma}$ is the ratio between viscous and capillary forces. The Reynolds number based on the shear rate and drop radius is $\mathrm{Re}\equiv\frac{\rho\dot{\gamma}R^2}{\mu}$.

The simulations were performed using the free-energy lattice Boltzmann method. With this diffuse interface method, topological changes of the interface do not need reconstruction of a mesh after breakup or coalescence of the droplets. However, a large number of lattice nodes is required to resolve both the drops and the thin film between them before they coalesce. To perform simulations with a sufficiently high resolution that allows us to capture the transition between coalescence and sliding after a collision, we use multiple GPUs in parallel. The simulations visualized in this video were performed using nine NVIDIA Tesla M2070 GPUs. The domain size was $1024 \times 512 \times 512$ nodes, and the initial radius of the spherical drop was 100 nodes.  

The simulations mimic experimental work on droplet collisions in shear flow. The first simulation involves drop breakup at $\mathrm{Ca}=0.2$ and $\mathrm{Re}=10$. The shear is stopped at $\dot{\gamma}t=22.2$, which is before the droplet breaks, to control the number of droplets that form. Due to inertia, the droplet continues to stretch and reaches a maximum elongation before shrinking back. As it retracts, a thin neck suddenly forms. Capillary forces dominate and the drop breaks into two daughter droplets. If the shear were stopped later, several additional satellite droplets would also form. The final horizontal droplet separation is $\frac{\Delta x}{2R}=1.77$ and the vertical offset is $\frac{\Delta y}{2R}=0.65$. Starting from this final state for the first simulation, two further numerical experiments were performed.

To observe collisions between the two droplets, we reverse the shear flow and consider two shear rates. The new capillary and Reynolds numbers were computed using the radius of the smaller droplets. In the first collision simulation, $\mathrm{Ca}=0.24$ and $\mathrm{Re}=9.6$. Under these conditions, the droplets slide without coalescing. Both physical and simulated droplets do not coalesce unless the capillary number is below a critical value that separates the regions of coalescence and non-coalescence. In the second case, we repeat the numerical experiment with a lower shear rate at which $\mathrm{Ca}=0.08$ and $\mathrm{Re}=3.2$. This time the drops coalesce because the collision is sufficiently slow to allow the film between them to drain.

\subsection*{Acknowledgment}
This research has been enabled by the use of computing resources provided by WestGrid and Compute/Calcul Canada.

\end{document}